\documentclass[preprint,showpacs,preprintnumbers,amsmath,amssymb,showkeys]{revtex4}
\usepackage{graphicx}
\usepackage{dcolumn}
\usepackage{bm}

\begin{document}
\renewcommand{\thesection}{\arabic{section}}
\renewcommand{\thetable}{\arabic{table}}
\setlength{\baselineskip}{16.0pt}

\bibliographystyle{unsrt}
\title{\center{ A note regarding Gram-Schmidt states on $T^2$}}

\author{Mario Encinosa }
\affiliation{ Florida A\&M University \\Department of Physics \\
205 Jones Hall \\ Tallahassee FL 32307}
\email{encinosa@cennas.nhmfl.gov}

\newpage


\begin{abstract}
An efficient  procedure for generating  Gram-Schmidt states on a
toroidal surface $T^2$ is presented. As an application of the
method, low-lying eigenvalues and wave functions for an electron
on $T^2$ subjected to a constant magnetic field are determined.
\end{abstract}

\pacs{03.65Ge,02.60.Cb }
\keywords{Gram-Schmidt, torus, magnetic field}
\maketitle

 Toroidial systems are relevant to fusion  \cite{kucinski}, heat
transfer \cite{ausder,chapko}, vibrational
\cite{inman,sarigul,madureira,zhou}, electromagnetic
\cite{shushkevic} and, recently, nanostructure physics
\cite{fuhrer,latge, lorke, garsia}. For problems restricted to the
neighborhood of a toroidal surface, i.e., to those on or near
toroidal shells, Gram-Schmidt (GS) functions on $T^2$ could
potentially prove useful. As  noted elsewhere, the calculation of
higher order GS states may be time consuming and oftentimes
subject to numerical error \cite{madureira} when performed in a
black-box fashion, particulary when trigonometric functions are
employed as the primitive basis set and integrations are performed
over a non-trivial integration measure. In this note we show that
there exists a simple algorithm for constructing higher order GS
functions on $T^2$ that eliminates these difficulties and proceed
to apply the functions to the problem of an electron on $T^2$ in a
magnetic field directed along the z-axis.

The geometry adopted here parameterizes a toroidal surface of
major radius $R$ and minor radius
 $a$ in terms of  cylindrical coordinate unit vectors by
\begin{equation}
 \mathbf{r} (\theta,\phi)=W (\theta){\bm {\rho}} +a\  {\rm sin}
\theta{\bm  {k}}  \end{equation} with $W=R + a \ {\rm cos} \theta
$.
 Applying $d$ to Eq.(1)
gives
\begin{equation}
d \mathbf{r}= a  d\theta \ {\bm \theta}+W d\phi{\bm \phi}
\end{equation}
with $\mathbf{\theta} =-\rm sin \theta  \mathbf{\rho}+\rm cos
\theta \mathbf{k}$.
 The metric elements $g_{ij}$ can be read off of
\begin{equation}
d{\bf r}\cdot d \mathbf{r}=a^2 d\theta^2+
                W^2d\phi^2
\end{equation}
so that the integration measure is
\begin{equation}
{\sqrt g}dq^1dq^2 \rightarrow  a W d\theta d\phi.
\end{equation}
It will prove useful in what follows to write the Laplacian
derived from Eq. (3); it is \cite{fpl}
\begin{equation}
\nabla^2 = {1 \over a^2}{\partial^2   \over \partial \theta^2} -
 { { \rm sin}\ \theta \over a W}
 {\partial   \over \partial
 \theta}
+{1 \over W^2}{\partial^2   \over
\partial \phi^2}.
\end{equation}

 The solutions of the Schrodinger equation derived
from Eq. (5) can be taken of the form $\psi_n(\theta)e^{i\nu\phi}$
with $\nu \equiv$ integer. A procedure for determining  free
particle surface wave functions $\psi_n(\theta)$ has been given
elsewhere \cite{fpl} and extended to several special cases in
\cite{halberg1,halberg2,halberg3}. However, if an arbitrary
potential $V(\theta,\phi)$ is included in the Hamiltonian, many
 $\theta, \phi$ states may be necessary in a basis set expansion to achieve suitable
convergence for the eigenvalues and wave functions of
$H(\theta,\phi)$.

The   $\theta \rightarrow -\theta$ symmetry of the Laplacian
allows the solutions of the Schrodinger equation  to be split into
even and odd functions, and the primitive basis set can be taken
to possess this property,
\begin {equation} u_n(\theta) = {1 \over \sqrt
\pi} {\rm cos}[n\theta], \qquad v_n(\theta) = {1 \over \sqrt \pi}
{\rm sin}[n\theta].\end{equation}

 Here for the sake of brevity, we will
consider only the even  functions in detail and later comment
briefly on the simple modification needed for the odd functions.
Setting $\alpha = a/R$ and $F(\theta)=1+\alpha \ \rm cos\theta$,
the simple but key point in what follows is the integrals
encountered in the GS procedure (irrelevant factors of $R$ and $a$
are dropped)
\begin{equation}
t_{n \bar{n}}=\int^{2\pi}_0
u_n(\theta)u_{\bar{n}}(\theta)F(\theta)d\theta
\end{equation}
are non-zero only for
\begin{equation}
t_{00}=2, \qquad t_{01}=\alpha;
\end{equation}
\begin{equation}
 t_{nn}=1, \ \qquad t_{n,n+1}=\alpha /2, \qquad n>0.
 \end{equation}

 Write
 \begin{equation}
\psi_n = \sum_{m=0}^n c_{nm} u_m,
\end{equation}
 and use Eqs. (7) - (9) in the standard GS procedure;  the
  first few GS states are sufficient to illustrate the general
pattern that emerges $(\beta = \alpha
 /2)$,
\begin{equation}
 \psi_0= u_0 / \sqrt 2  \equiv N_0 u_0
 \end{equation}
 \begin{equation}
 \psi_1=N_1[u_1-N_0 \beta N_0u_0]
 \end{equation}
 \begin{equation}
 \psi_2=N_2[u_2 - N_1 \beta N_1u_1 + N_1 \beta N_1 N_0 \beta N_0u_0]
\end{equation}
 \begin{equation}
 \psi_3=N_3[u_3 - N_2 \beta N_2 u_2+ N_2 \beta N_2
 N_1 \beta N_1 u_1 - N_2 \beta N_2 N_1 \beta N_1 N_0 \beta N_0u_0]
\end{equation}
{\centerline \vdots}
from which it is apparent that
\begin{equation}
c_{nm}=(-)^{n+m}N_n(N_{n-1}\beta N_{n-1})(N_{n-2}\beta
N_{n-2})...(N_{m}\beta N_{m}) .
\end{equation}
To obtain the $c_{nm}$ the normalization factors must be
determined. Consider the $k^{th}> 1$ unnormalized state $\Phi_k$;
it is easy to establish $N_k$ from
\begin{equation}
\ < \Phi_k |  \ \Phi_k   > =  1 - N_{k-1}\beta N_{k-1}
\end{equation}
since all $< u_i | u_{i+j} >$ vanish for $j > 1$. Starting then
from $N_1^2 = (1-2\beta^2)^{-1}$, for $k > 1$ the squared
normalization for a given $\Phi_k$ is
\begin{equation}
N^2_{k+1}={1 \over {1-\beta^2 N^2_{k}}}.
\end{equation}
In summary, Eq. (17) serves to generate all normalization factors
and Eq. (15) the GS coefficients. The sinusoidal function
coefficients may be obtained by the method employed above by
starting with $c_{11}$ rather than $c_{00}$ and letting $N_1 = 1$.

It is worth noting that from Eq. (15) the ratio of successive
coefficients within a given $(n,m)$  series
\begin{equation}
c_{n,m}/c_{n,m+1} =- N_m \beta N_m
\end{equation}
 allows for
 truncation approximations to be made for certain problems. To
illustrate this point, write Eq. (17) in continued fraction form;
taking for example $N_4$ (which is the largest $N_k$ that can be
comfortably typeset),
\begin{equation}
 N^2_4 = \
    \frac{1}{1-\frac{\beta^2}{1-\frac{\beta^2}{1-\frac{\beta^2}{1-2\beta^2{}}}}} \ \ .
\end{equation}
Any symbolic algebra program can be utilized to evaluate this
expression, but once known to a given order of $\alpha = 2\beta$
it need not be evaluated any further. To $O(\alpha^8)$ which
occurs at $N_5$,
\begin {equation}
N^2_5 = 1 + {\alpha^2 \over 4}+{\alpha^4 \over 8}+{5\alpha^6 \over
64}+{7\alpha^8 \over 128}+O(\alpha^{10})+...,
\end{equation}
and a final ${\alpha  \over 2}$ factor multiplies the expression
of Eq. (20) to establish right hand side of Eq. (18). Since
$\alpha < 1$,  concatenating several $N \beta N$ factors causes
the series to converge rapidly for larger $n$ and smaller $m$. For
quantum mechanical applications, the kinetic energy operator
yields an $m^2$ factor for each term in the expansion given by Eq.
(10), making it possible to truncate higher states at say, three
terms for sufficiently large $n$ (as set by the particular
problem)
\begin{equation}
\psi_n(\theta) =  c_{nn} \ {\rm cos} \ n\theta + c_{n,n-1}\ {\rm
cos} \ (n-1)\theta + c_{n,n-2}\ {\rm cos} \ (n-2)\theta.
\end{equation}

An obvious immediate  application for the GS states of relevance
to nanoscience \cite{lorke,garsia,fuhrer} is the problem of an
electron on  $T^2$ in the constant magnetic field
\begin{equation}
\mathbf{B} = B_0 \mathbf{k}.
\end{equation}
In the Coulomb gauge $\nabla \cdot \mathbf {A} = 0$ the vector
potential ${\mathbf A}(\theta,\phi) = {1 \over 2} \mathbf{B}
\times \mathbf{r} $ as expressed in surface variables is
\begin{equation}
\mathbf {A}(\theta,\phi) = {B_0 R F \over 2} \bm {\phi}.
\end{equation}
 The Schrodinger
equation (here the spin splitting is neglected)
\begin{equation}
 H = {1 \over {2m}}\bigg ( {\hbar
\over i} \nabla + q {\mathbf A} \bigg) ^2\Psi = E\Psi
\end{equation}
is more simply expressed if we first define
$$  \gamma_0 = B_0 \pi R^2 $$
$$ \gamma_N = {\pi  \hbar \over q} $$
$$ \tau_0 = {\gamma_0 \over \gamma_N}$$
$$ \varepsilon = {2m_e Ea^2 \over \hbar^2}$$
after which Eq. (24) may be put in the form
\begin{equation}
 \bigg ( {\partial^2 \over \partial^2 \theta} -
  {\alpha \  {\rm sin} \ \theta \over F}{\partial \over \partial
 \theta} + {\alpha^2 \over F^2}{\partial^2 \over
\partial^2 \phi} +
 i \tau_0\alpha^2{\partial \over \partial
 \phi}
-{\tau_0^2 \alpha^2F^2 \over 4}+\varepsilon
  \bigg)\Psi
\end{equation}
\begin{equation}
\equiv (H_\tau\ + \varepsilon) \Psi = 0.
\end{equation}

The basis expansion functions are taken as per Eqs. (11) - (14)
with azimuthal eigenfunctions appended (the magnetic field term
considered here does not cause even (+) and odd (-)  functions to
mix),
\begin{equation}
\psi^{\pm}_{n\nu}(\theta,\phi) = {1 \over \sqrt {2
\pi}}\sum_{m}c^{\pm}_{nm} \left ( \begin{array}{c} u_m(\theta) \\
v_m(\theta) \end{array} \right)
 e^{i\nu\phi}.
\end{equation}
The matrix
\begin{equation}
H_{\tau n \bar{n}} = \big < \bar{n}|H_\tau|n \big >
\end{equation}
is then easily constructed since the matrix elements can all be
written in closed form, and the eigenvalues and eigenvectors
determined with a six-state expansion for each $\theta$-parity
\cite{mat} (no truncations were performed).

 Table 1 gives the even function
ground and first excited state energy eigenvalues and wave
functions for several values of $\nu$ and $\tau_0$.  Table 2 gives
the same for the odd functions \cite{parnote}. Three decimal place
accuracy was achieved for the eigenvalues/  wave functions for
four of the six states generated from $H_{\tau n \bar{n}}$ when
$\tau_0 = 10$, which corresponds to a field of $2.6  \ T$ for a
torus with $R = 50 \ nm$. Although a large scale treatment of this
problem was considered outside the scope of this work, it should
be noted that because the basis states are so simple to generate
the only inherent limitation to such a treatment is the matrix
inversion. Additionally, the matrix elements of the Hamiltonian
for an off-axis magnetic field that comprises the general case
\begin{equation}
{\mathbf B} = B_x{\mathbf i} + B_z{\mathbf k}
\end{equation}
 can also be done in closed form given sufficient patience \cite{mottpc}.

 In conclusion, we have presented an algebraic method to derive
 GS states on $T^2$ with very little effort. The ability to obtain these
 functions rapidly may
prove of use to problems in the areas mentioned in the opening
paragraph of this note. As an example of their utility, we have
used them to calculate the spectra of an electron on the surface
of a torus in the presence of a magnetic field. \vskip 16pt

\begin{center} {\bf Acknowledgments} \end{center} The author would like to thank B.
Etemadi for his encouragement and support.


\begin{thebibliography}{10}
\bibitem{parnote}
Parity here has only referred to the evenness/oddness of the basis
functions;
  in general the quantum mechanical parity will depend on the azimuthal~quantum
  number.
\end{thebibliography}

\newpage

\begin{table}
\caption{Even function (the (+) superscript has been supressed)
ground and first excited state energies and wave functions.  GS
state coefficients not shown in the table are much smaller than
those given. $-\nu$ states are trivially obtained from those given
below.}
\begin{center}
\begin{tabular}{|c|c|c|c|c|c|}
\hline $\nu$ & $\tau_0$ &$\varepsilon_0$ & $\Psi_0$ &
$\varepsilon_1$
& $\Psi_1 $\\
\hline

$\ 0$ & \ 0 & 0 & $\psi_0$ & 1.122 & $.997\psi_1-.082\psi_2+.014\psi_3$ \\
$\ 0$ & \ 5  & .139  & $.840\psi_0-.533\psi_1+.099\psi_2$ & 3.208 & $.531\psi_0+.772\psi_1-.345\psi_2$\\
$\ 0$ & \ 10 & -.747 & $.744\psi_0-.637\psi_1+.198\psi_2$ & 7.925 & $.466\psi_0+.289\psi_1-.751\psi_2$  \\
\hline
$1$ & 0 & .249 & $-.987\psi_0-.162\psi_1+.021\psi_2$ & 1.1663 & $-.162\psi_0+.987\psi_1+.006\psi_2$ \\
$1$ & 5  & 1.955  & $.882\psi_0-.469\psi_1+.052\psi_2$ & 4.806 & $.462\psi_0+.838\psi_1-.289\psi_2$\\
$1$ & 10 & 2.429 & $.759\psi_0-.627\psi_1+.176\psi_2$ & 10.927 & -$.457\psi_0-.327\psi_1-.755\psi_2-.333\psi_3$  \\
\hline
$2$ & 0 & .795 & $.931\psi_0+.364\psi_1-.018\psi_2$ & 3.175 & $-.342\psi_0+.891\psi_1+.296\psi_2$ \\
$2$ & 5  & 4.569  & $.973\psi_0-.221\psi_1-.068\psi_2$ & 7.198 & $.217\psi_0+.973\psi_1-.063\psi_2$\\
$2$ & 10 & 6.851 & $-.803\psi_0+.587\psi_1-.105\psi_2$ & 14.940 & $.423\psi_0+.448\psi_1-.7747\psi_2+.247\psi_3$  \\
 \hline
\end{tabular}
\end{center}
\end{table}

\begin{table}
\caption{Odd function (the (-) superscript has been supressed)
ground and first excited state energies and wave functions.  GS
state coefficients not shown in the table are much smaller than
those given. $-\nu$ states are trivially obtained from those given
below.}
\begin{center}
\begin{tabular}{|c|c|c|c|c|c|}
\hline $\nu$ & $\tau_0$ &$\varepsilon_0$ & $\Psi_0$ &
$\varepsilon_1$
& $\Psi_1 $\\
\hline

$\ 0$ & \ 0 & .977 & $.996\psi_1-.086\psi_2$ & 4.033 & $-.087\psi_1-.991\psi_2+.103\psi_3$ \\
$\ 0$ & \ 5  & 2.510  & $.948\psi_1-.316\psi_2+.047\psi_3$ & 5.834 & $-.317\psi_1-.914\psi_1+.250\psi_3$\\
$\ 0$ & \ 10 & 5.749 & $.757\psi_1-.617\psi_2+.210\psi_3$ & 11.128 & $.600\psi_1+.530\psi_2-.569\psi_3$  \\
\hline
$1$ & 0 & 1.264 & $.999\psi_1-.037\psi_1$ & 4.411 & $-.036\psi_1-.997\psi_2+.067\psi_3$ \\
$1$ & 5 & 4.125  & $.963\psi_1-.269\psi_2+.026\psi_3$ & 7.441 & $-.268\psi_1-.938\psi_2+.218\psi_3$\\
$1$ & 10& 8.760 & $.776\psi_1-.602\psi_2+.186\psi_3$ & 14.000 & -$.581\psi_1-.567\psi_2+.558\psi_3-.166\psi_4$  \\
\hline
$2$ & 0 & 2.041 & $-.995\psi_1-.089\psi_2$ & 5.532 & $.088\psi_1-.995\psi_2-.046\psi_3$ \\
$2$ & 5  & 6.359  & $.991\psi_1-.132\psi_2-.022\psi_3$ & 9.774 & $.129\psi_1+.985\psi_2-.113\psi_3$\\
$2$ & 10 & 12.710 & $-.829\psi_1+.548\psi_2-.115\psi_3$ & 17.611 & $.518\psi_1+.676\psi_2-.513\psi_3+.107\psi_4$  \\
 \hline
\end{tabular}
\end{center}
\end{table}


\end{document}